\documentclass[seceq,supplement]{ptptex}
\usepackage{amssymb,amsmath,graphicx,epsfig,wrapfig}

\title{Determination of fragmentation functions \\
       and their application to exotic-hadron search}

\author{M. Hirai$^{\, a}$ and S. Kumano$^{\, b}$}

\inst{
$^a$ Department of Physics, 
    Tokyo University of Science \\ 
    2641, Yamazaki, Noda, Chiba, 278-8510, Japan \\
$^b$ KEK Theory Center,
    Institute of Particle and Nuclear Studies, KEK \\
    and Department of Particle and Nuclear Studies \\
    Graduate University for Advanced Studies \\
    1-1, Ooho, Tsukuba, Ibaraki, 305-0801, Japan
}

\abst{
We discuss studies on determination of fragmentation functions
and an application to exotic-hadron search by using characteristic
differences between favored and disfavored functions.
The optimum fragmentation functions are determined 
for pion, kaon, and proton in the leading order (LO) 
and next-to-leading order (NLO) of the running coupling 
constant $\alpha_s$ by global analyses 
of hadron-production data in electron-positron annihilation.
Various parametrization results are much different in disfavored-quark
and gluon fragmentation functions; however, we show that they are 
within uncertainties of the determined functions by using 
the Hessian method for uncertainty estimation. We find that 
the uncertainties are especially large in the disfavored-quark 
and gluon fragmentation functions. NLO improvements are explicitly
shown in the determination by comparing uncertainties of 
the LO and NLO functions.
Next, we propose to use differences between favored and disfavored 
fragmentation functions for determining internal quark configurations
of exotic hadrons. We make a global analysis for $f_0 (980)$ for
finding its internal configuration; however, uncertainties are
too large to specify the structure at this stage.
}

\begin{document}

\maketitle

\section{Introduction}
\label{intro}

Semi-inclusive hadron-production processes are important 
for investigating properties of quark-hadron matters in 
heavy-ion collisions and for finding the origin of the nucleon spin
in lepton-nucleon scattering and polarized proton-proton collisions.
Fragmentation functions (FFs) are key quantities in describing such
hadron-production processes in high-energy reactions. 
They indicate hadron-production probabilities from partons.

The FFs have been determined mainly by hadron-production data
of $e^+ e^-$ reaction. Many measurements were done in the $Z^0$-mass
region, whereas lower-energy data are not sufficient.
The determination of the FFs is not in an excellent situation in comparison
with the one of parton distribution functions (PDFs), which is obvious
from the fact that there are huge differences between the FFs of
different analysis groups, especially for disfavored-quark and
gluon FFs \cite{ffs_before_2006}. It led us to investigate
uncertainties of the FFs \cite{hkns07} as it was done
in the PDFs for the nucleon and nuclei \cite{pdf-errors,errors}.
In addition, making global analyses in the leading order (LO) of the
running coupling constant $\alpha_s$ and the next-to-leading 
order (NLO) at the same time, we showed that the role of
NLO terms for reducing the uncertainties of the determined
FFs. We provided a useful code in calculating the optimum
FFs for the pion, kaon, and proton from our global analyses
\cite{hkns07,ffs-web}. After our studies, there are works 
on related global analyses of the FFs \cite{ffs-summary, ffs-recent}
and also on a hadron-model estimate \cite{ffs-njl}.

Next, we proposed a possible method for exotic-hadron search
by using the FFs \cite{hkos08}. In particular, the FFs are
usually classified into favored and disfavored functions.
The favored means the fragmentation from a quark which exists 
in a hadron $h$ as a constituent in a naive quark model.
The disfavored means the fragmentation from a sea quark.
Therefore, internal quark configuration should be reflected
in both FFs. This fact led us to investigate an interesting
suggestion to use the FFs for a possible exotic-hadron search
by looking at differences between the favored and disfavored
functions \cite{hkos08}.

We explain these works in this article.
In Sec. \ref{ffs}, the FFs are defined in $e^+ e^-$ annihilation
processes. Our global analysis method is explained in Sec. \ref{method},
and results are shown in Sec. \ref{results}. 
The idea of using the FFs for exotic-hadron search is introduced
in Sec. \ref{exotic}. Our studies are summarized in Sec. \ref{summary}.

\section{Fragmentation functions in $e^+ e^-$ annihilation}
\label{ffs}

The cross section of hadron-$h$ production in the $e^+ e^-$ annihilation
is described by a $q\bar q$-pair production
$e^+ e^- \rightarrow q\bar q$ followed by a fragmentation process
from $q$ ($\bar q$ or gluon emitted from $q$ or $\bar q$) to 
the hadron $h$. The FF is defined by the cross section
\begin{equation}  
F^h(z,Q^2) = \frac{1}{\sigma_{tot}} 
\frac{d\sigma (e^+e^- \rightarrow hX)}{dz} ,
\label{eqn:def-ff}
\end{equation}
where $\sigma_{tot}$ is the total hadronic cross section, and
$Q^2$ is the virtual photon or $Z^0$ momentum squared
in $e^+e^- \rightarrow \gamma, Z^0$. It is equal to the center-of-mass
energy squared $s$ ($=Q^2$). The variable $z$ is defined by the energy
fraction: 
\begin{equation}   
z \equiv \frac{E_h}{\sqrt{s}/2} = \frac{2E_h}{Q},
\label{eqn:def-z}
\end{equation}
where $E_h$ is the hadron energy. Namely, $z$ is the hadron energy
scaled to the beam energy ($\sqrt{s}/2$). The fragmentation is
described by the summation of all the parton contributions:
\begin{equation}  
F^h(z,Q^2) = \sum_i C_i(z,\alpha_s) \otimes D_i^h (z,Q^2).
\label{eqn:def-ffqqbarg}
\end{equation}
Here, $D_i^h(z,Q^2)$ is a fragmentation function of the hadron $h$ 
from a parton $i$ ($=g, \ u,\ d,\ s,\ \cdot\cdot\cdot$),
$C_i(z,\alpha_s)$ is a coefficient function which is calculated
in perturbative QCD \cite{ffs_before_2006,qqbar-cross}, and
the convolution integral $\otimes$ is defined by
$f (z) \otimes g (z) = \int^{1}_{z} dy / y \,  f(y) g (z/y) $.

The measurements of the FFs have been done in various $Q^2$,
whereas they are parametrized at a fixed $Q^2$ point ($\equiv Q_0^2$)
as explained in the next section. The initial functions at $Q_0^2$
are evolved to the experimental $Q^2$ points by the standard DGLAP 
evolution equations. The equations are essentially the same as the ones 
for the PDFs by exchanging the splitting functions $P_{qg}$ and $P_{gq}$. 
There are also some differences between their NLO expressions 
\cite{splitting}, for example, in the modified minimal subtraction 
($\overline {\rm MS}$) scheme. 

\section{Global analysis method}
\label{method}

The FFs are expressed in terms of a number of parameters, which are
determined by a $\chi^2$ analysis of the $e^+ + e^- \rightarrow h+X$
data. The initial functions are provided at $Q_0^2$ as
\begin{equation}
D_i^h(z,Q_0^2) = N_i^h z^{\alpha_i^h} (1-z)^{\beta_i^h} ,
\end{equation}
where $N_i^h$, $\alpha_i^h$, and $\beta_i^h$ are the parameters.
An apparent constraint for the parameters is the energy sum rule:
\begin{equation}
\sum_h M_i^h \equiv \sum_h \int_0^1 dz \, z \, D_i^h (z,Q^2) = 1 ,
\label{eqn:sum}
\end{equation}
where $M_i^h$ is the second moment of $D_i^h (z,Q^2)$.
However, it is almost impossible to confirm this sum since
the summation over all the hadrons cannot be taken practically.
In our analysis, we tried to be careful that the sum does not
significantly exceed 1 even within analyzed hadrons.

\begin{wraptable}{r}{0.50\textwidth}
\vspace{-0.4cm}
\caption{Experiments, center-of-mass energies,
     and numbers of data points are listed for used data sets
     of $e^+ +e^- \rightarrow \pi^\pm +X$ \cite{hkns07}.}
\label{tab:exp-pion}
\vspace{-0.2cm}
\begin{center}
\begin{tabular}{lcc} \hline \hline
Experiment           & $\sqrt{s}$ (GeV)        & \# of data \\
\hline
TASSO                & 12,14,22,30,34,44       &  29    \\
TPC                  & 29                      &  18    \\
HRS                  & 29                      & \, 2   \\
TOPAZ                & 58                      & \, 4   \\
SLD                  & 91.28                   &  29    \\
SLD (u,d,s)          & 91.28                   &  29    \\
SLD (c)              & 91.28                   &  29    \\
SLD (b)              & 91.28                   &  29    \\
ALEPH                & 91.2  \,                &  22    \\
OPAL                 & 91.2  \,                &  22    \\
DELPHI               & 91.2  \,                &  17    \\
DELPHI (u,d,s)       & 91.2  \,                &  17    \\
DELPHI (b)           & 91.2  \,                &  17    \\
\hline
Total                &                         & 264 \, \\
\hline
\end{tabular}
\end{center}
\end{wraptable}

There are two types in the FFs: favored and disfavored functions.
For the FFs of light quarks ($u,d,s$), they are assumed to be equal
if they are favored or disfavored ones. The favored functions are
given by
\begin{align}
D_{u}^{\pi^+} (z, & Q_0^2) = D_{\bar d}^{\pi^+} (z,Q_0^2)
\nonumber \\
          &   = N_{u}^{\pi^+} z^{\alpha_{u}^{\pi^+}}
                (1-z)^{\beta_{u}^{\pi^+}} ,
\label{eqn:favored}
\end{align}
for $\pi^+$.
The $\pi^+$ productions from $\bar u$, $d$, $s$, and $\bar s$ are
disfavored processes so that they are assumed to be equal
at $Q_0^2$:
\begin{align}
D_{\bar u}^{\pi^+} (z, & Q_0^2) = D_{d}^{\pi^+} (z,Q_0^2)
\nonumber \\
& = D_{s}^{\pi^+} (z,Q_0^2)
  = D_{\bar s}^{\pi^+} (z,Q_0^2)
\nonumber \\
    & = N_{\bar u}^{\pi^+} z^{\alpha_{\bar u}^{\pi^+}}
         (1-z)^{\beta_{\bar u}^{\pi^+}} .
\label{eqn:disfavored}
\end{align}
The FFs from a gluon and heavy quarks are defined separately as
\begin{align}
D_{g}^{\pi^+} (z,Q_0^2) 
& = N_{g}^{\pi^+} z^{\alpha_{g}^{\pi^+}} (1-z)^{\beta_{g}^{\pi^+}} ,
\nonumber \\ 
D_{c}^{\pi^+} (z,m_c^2) & = D_{\bar c}^{\pi^+} (z,m_c^2)
 = N_{c}^{\pi^+} z^{\alpha_{c}^{\pi^+}} (1-z)^{\beta_{c}^{\pi^+}} ,
\nonumber \\
D_{b}^{\pi^+} (z,m_b^2) & = D_{\bar b}^{\pi^+} (z,m_b^2)
 = N_{b}^{\pi^+} z^{\alpha_{b}^{\pi^+}} (1-z)^{\beta_{b}^{\pi^+}} ,
\label{eqn:gcb}
\end{align}
where $m_c$ and $m_b$ are charm- and bottom-quark masses.
The parameters in Eqs. (\ref{eqn:favored}), (\ref{eqn:disfavored}), 
and (\ref{eqn:gcb}) are determined so as to fit the data
in Table \ref{tab:exp-pion}, where experimental collaborations, 
center-of-mass energies, and the numbers of data are listed
for the charged-pion production.
It is clear that most data are taken at the $Z^0$ mass.

The FFs for the kaon and proton are parametrized in the similar way
by considering favored and disfavored functions. The amounts of
data are almost the same as the ones in Table \ref{tab:exp-pion}.
The detailed should be found in the original article \cite{hkns07}.
The parameters for the kaon and proton are determined
in separate $\chi^2$ analyses. 

One of our major purposes is to show the uncertainties of the FFs 
as explained in Sec. \ref{intro}. The uncertainties 
have been already estimated in the studies of nucleonic and nuclear PDFs
\cite{pdf-errors, errors}. The same Hessian method is used
for the uncertainty estimation. The $\chi^2$ is expanded around 
the minimum $\chi^2$ point $\hat \xi$:
$
 \Delta \chi^2 (\xi) = \chi^2(\hat{\xi}+\delta \xi)-\chi^2(\hat{\xi})
        =\sum_{i,j} H_{ij}\delta \xi_i \delta \xi_j \ ,
$
where $H_{ij}$ is called Hessian which is the second derivative
matrix, $\xi$ indicates a parameter set, and $\hat \xi$ is the set
at the minimum $\chi^2$ point.
The confidence region is given in the parameter space by
supplying a value of $\Delta \chi^2$. Using the Hessian matrix
obtained in a $\chi^2$ analysis, we estimated the uncertainty
of the FF by
$
[\delta D_i^h (z)]^2=\Delta \chi^2 \sum_{j,k}
\left[ \partial D_i^h (z) / \partial \xi_j  \right]_{\hat\xi}
H_{jk}^{-1}
\left[ \partial D_i^h (z) / \partial \xi_k  \right]_{\hat\xi} .
$
There are some variations among groups on the appropriate 
$\Delta \chi^2$ value for showing the uncertainty range
in a global analysis. The details are explained in our article \cite{hkns07} 
about our $\Delta \chi^2$ choice.

\section{Determined fragmentation functions for pion, kaon, and proton}
\label{results}

\begin{wrapfigure}{r}{0.42\textwidth}
   \vspace{-0.2cm}
   \begin{center}
       \epsfig{file=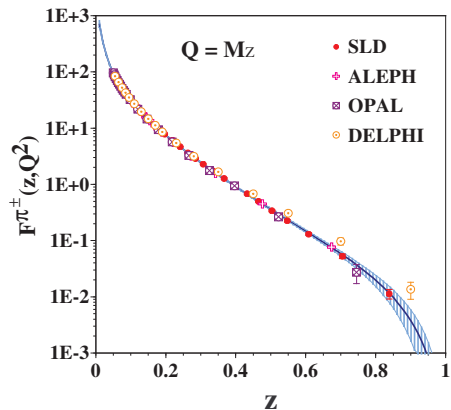,width=0.42\textwidth} \\
   \end{center}
   \vspace{-0.2cm}
\caption{Charged-pion fragmentation function in $e^+e^-$ 
         annihilation process. Our analysis results are 
         compared with experimental data \cite{hkns07}.}
\label{fig:pion-q-data}
\end{wrapfigure}
\vspace{-0.3cm}

\begin{figure}[t!]
   \vspace{-0.0cm}
   \begin{center}
       \epsfig{file=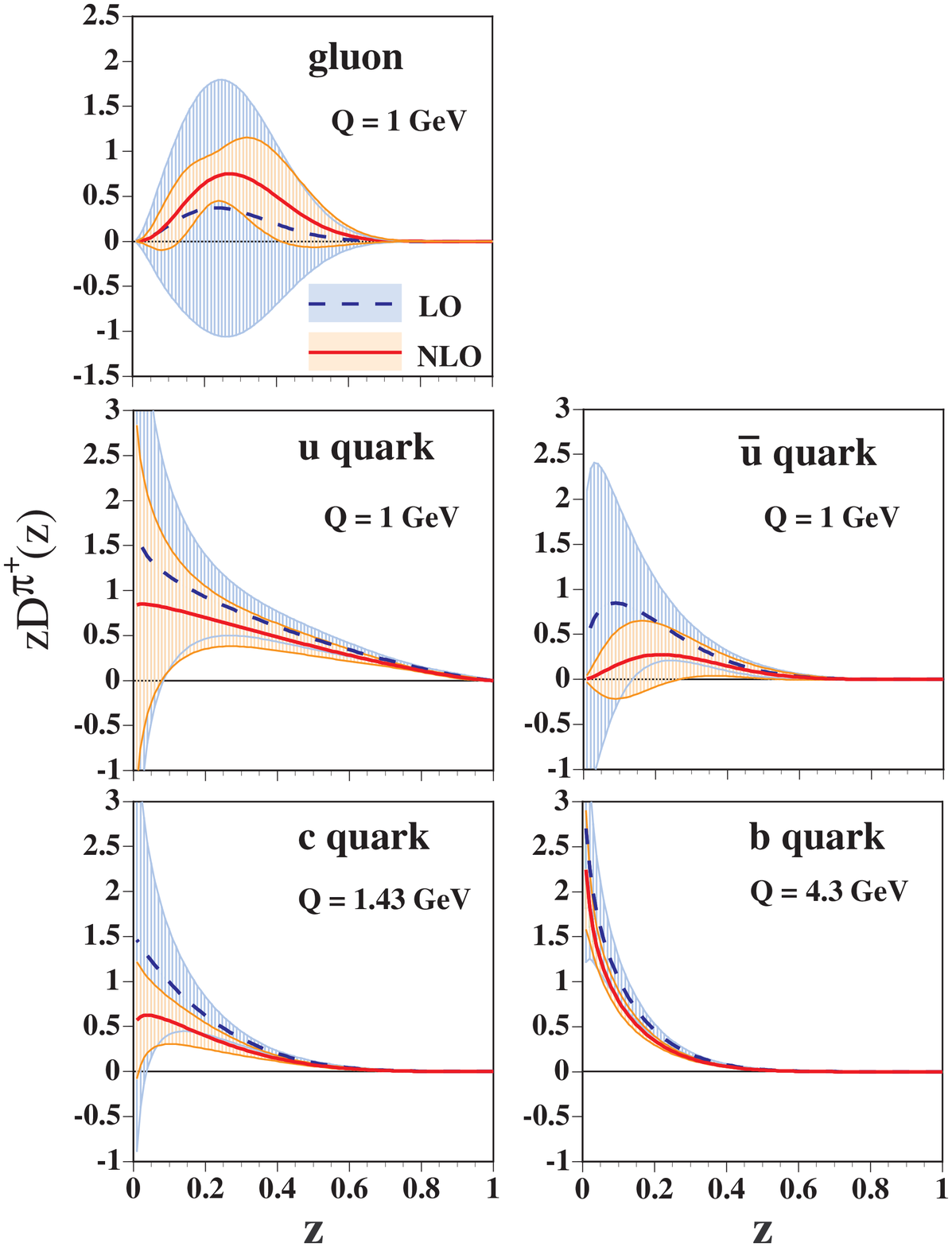,width=0.40\textwidth} 
       \hspace{0.5cm}
       \epsfig{file=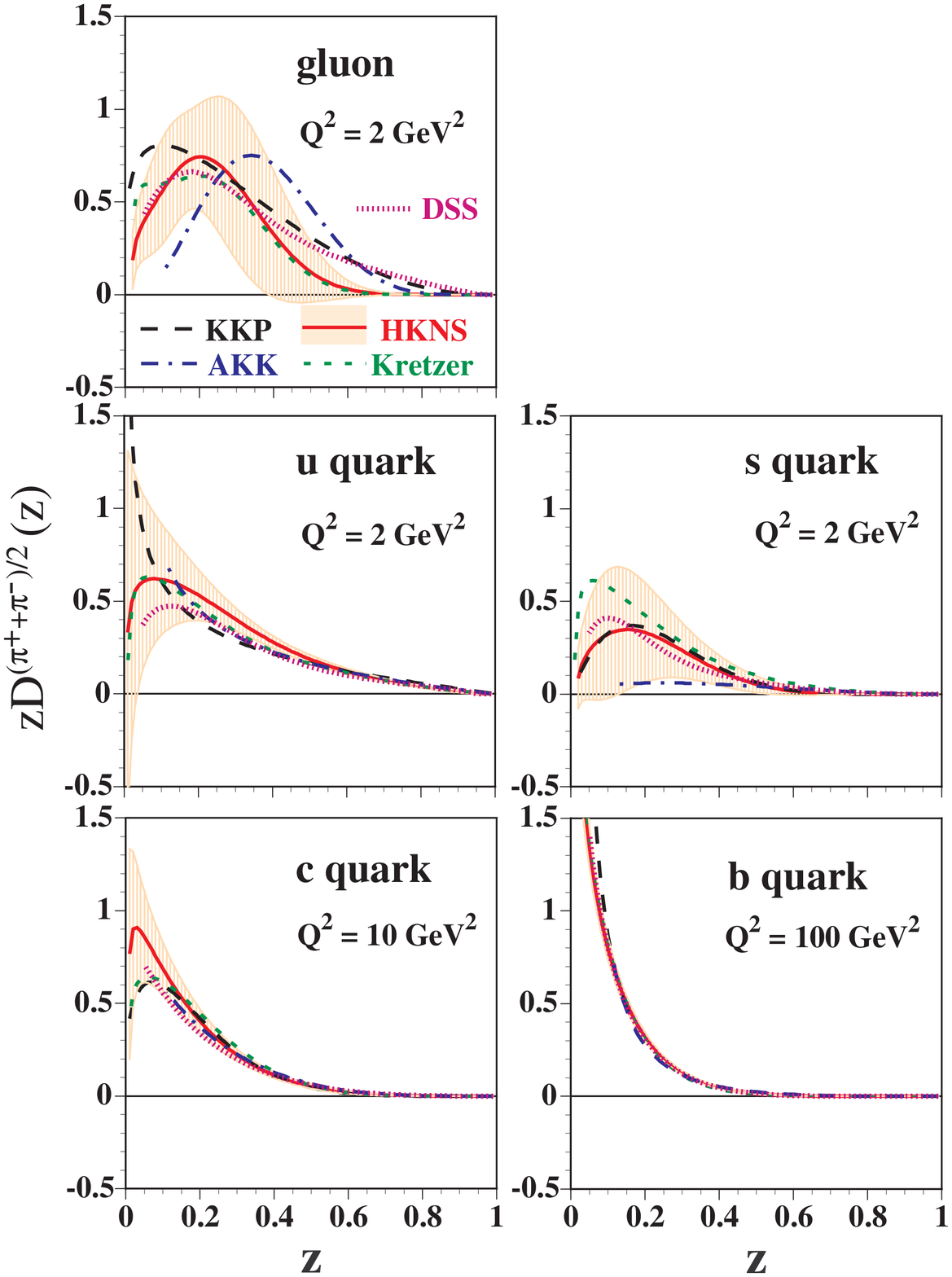,width=0.40\textwidth} \\
   \end{center}
   \vspace{-0.1cm}
\caption{Determined fragmentation functions for the pion 
         and their comparison with other parametrizations 
         \cite{hkns07, inpc07}. The shaded bands indicate
          estimated uncertainties.}
\label{fig:pion-ffs-comp}
   \vspace{0.5cm}
   \begin{center}
       \epsfig{file=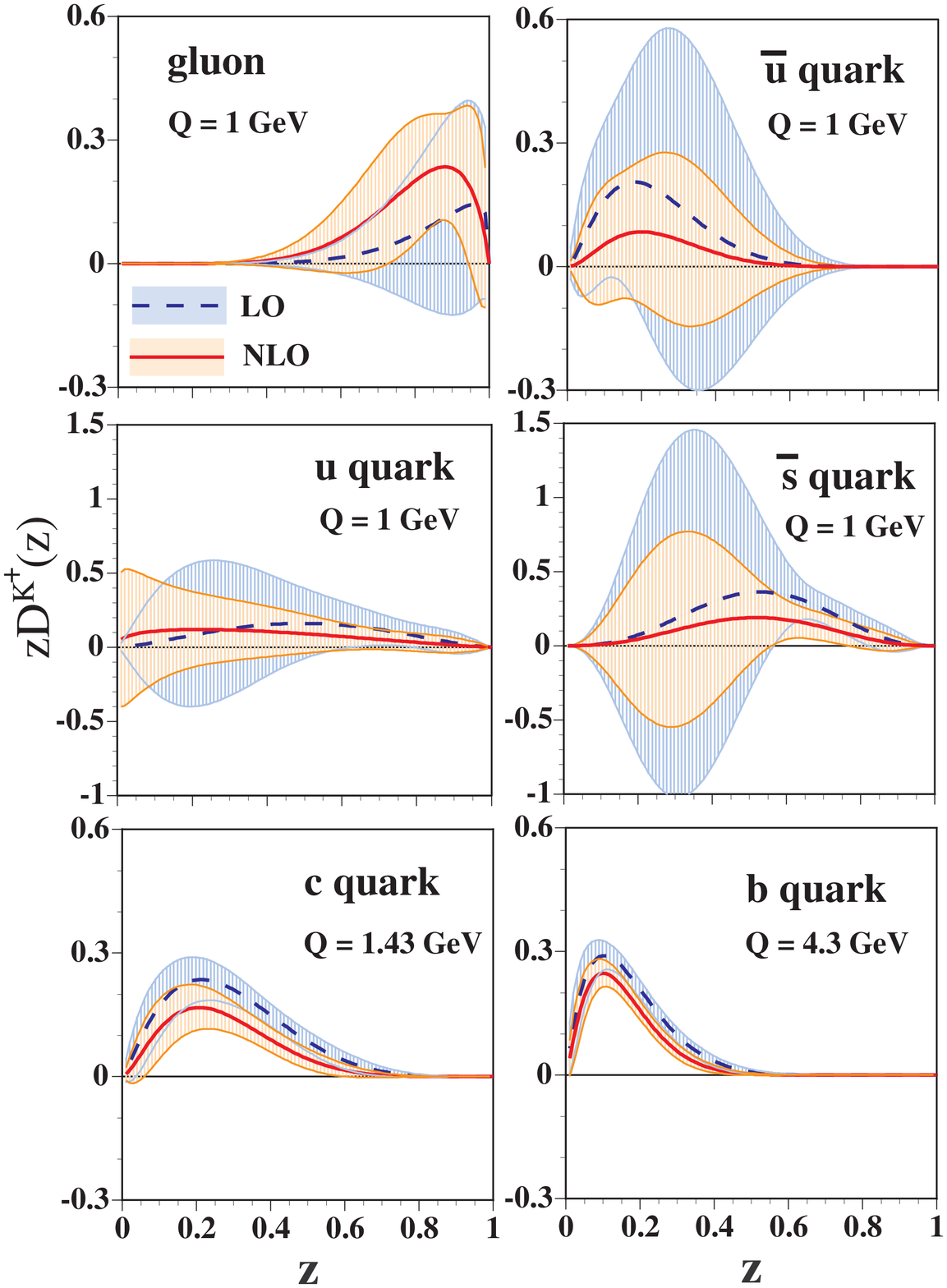,width=0.40\textwidth} 
       \hspace{0.5cm}
       \epsfig{file=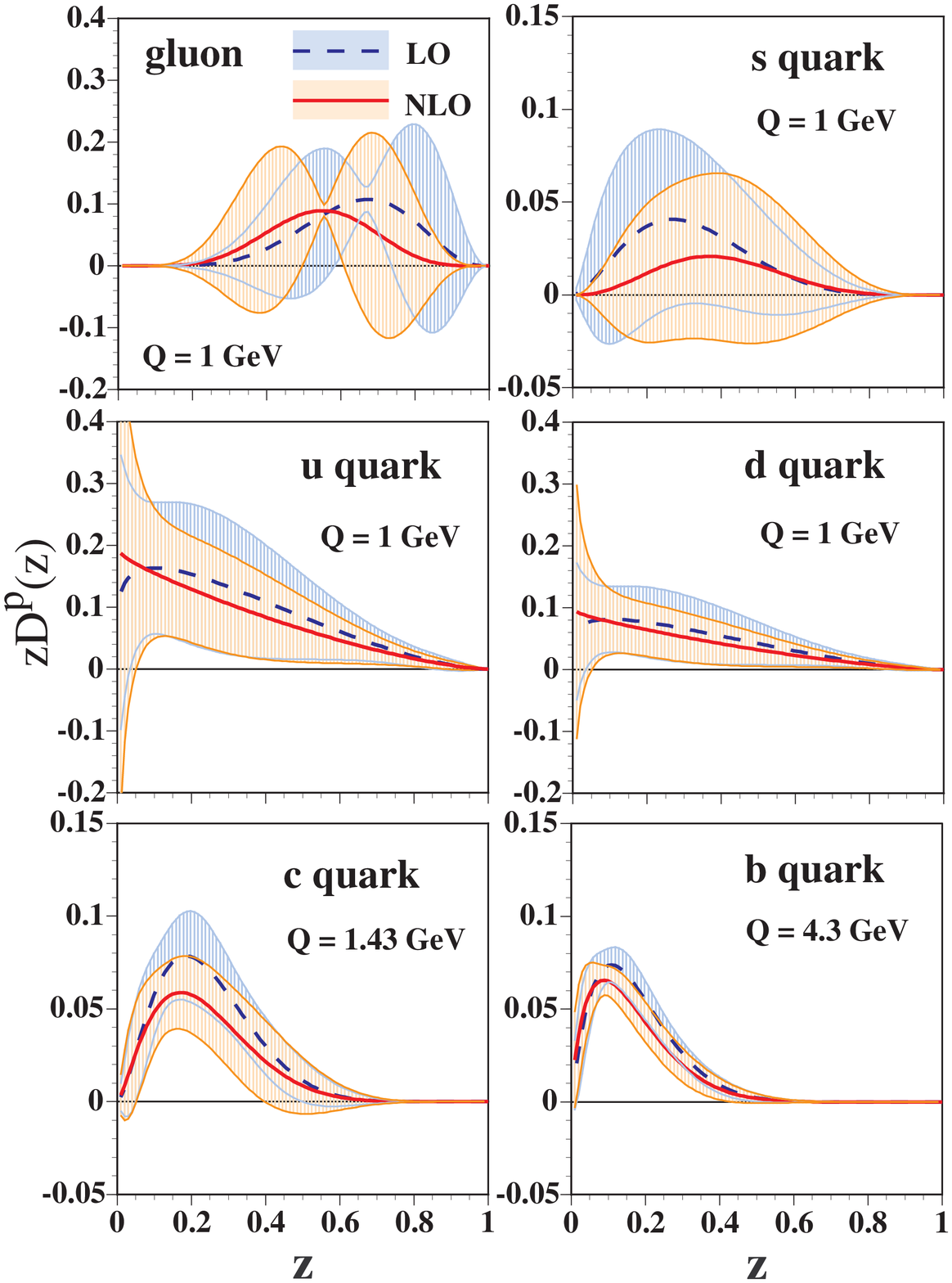,width=0.40\textwidth} \\
   \end{center}
   \vspace{-0.1cm}
\caption{Determined fragmentation functions for the kaon and 
         proton \cite{hkns07}.}
\label{fig:kp-ff-comp}
\end{figure}

We show obtained FFs of the pion by the $\chi^2$ analyses 
of the $e^+ e^- \rightarrow \pi^{\pm} X$ data
in Fig. \ref{fig:pion-q-data}, where the FF data in the form of
Eq. (\ref{eqn:def-ff}) and our parametrization result 
with an uncertainty band are shown at $Q^2=M_Z^2$.
The good agreement with the data indicates that the fit is 
successful from small- to large-$z$ regions.

Next, each FF is shown for the pion on the left-hand-side of
Fig. \ref{fig:pion-ffs-comp} with uncertainty bands in both LO and 
NLO ($\overline {\rm MS}$). It should be noted that the charm- and
bottom-quark FFs are shown at the scale of their mass thresholds 
$Q^2$=$m_c^2$ or $m_b^2$, whereas the others are shown at $Q^2$=1 GeV$^2$.
The uncertainties are generally larger in the LO, which indicates 
the FF determination is improved due to NLO terms.
We notice that disfavored-quark and gluon FFs 
have large uncertainties in both LO and NLO.

The NLO functions are then compared with other parametrizations 
of KKP, Kretzer, AKK, and DSS in Fig. \ref{fig:pion-ffs-comp}.
Some disfavored-quark and gluon functions, for example $s$-quark 
functions of Kretzer and AKK, are completely different
between the analysis groups; however,
they are within our uncertainty bands. 
There are not much differences between
the groups in favored- ($u$), charm-, and bottom-quark functions 
except for the small-$z$ region.
The large discrepancies among the different parametrizations 
especially in the disfavored-quark and gluon functions 
were not clearly understood before our work. 
Therefore, it is important to point out in our studies that
they are not due to an inappropriate analysis of some groups
and that they come from experimental errors and inaccurate
flavor decomposition for light quarks. It is simply impossible
to determine accurate disfavored-quark and gluon FFs from current
experimental data.

For the kaon and proton, our LO and NLO FFs are shown 
in Fig. \ref{fig:kp-ff-comp}. We found that both FFs are not
better determined than the pion FFs. There is a similar tendency 
with the pion case that the disfavored-quark and gluon FFs are not 
determined well. The NLO improvement, namely the uncertainty reduction,
is clear in the kaon, whereas it is not apparent in the proton.
We also found \cite{hkns07} in both kaon and proton that 
all the different analyses are consistent with each other 
in spite of their large differences in some functions
because they are roughly within our uncertainty bands. 

\section{Exotic-hadron search by fragmentation functions}
\label{exotic}

\begin{wraptable}{r}{0.50\textwidth}
\vspace{-0.3cm}
\caption{Second moments of the FFs in the NLO for $\pi^+$, $K^+$, and $p$
         at $Q^2$=1 GeV$^2$.}
\label{tab:2nd-moments}
\vspace{-0.0cm}
\begin{center}
\begin{tabular}{lcc} \hline \hline
Type             & Function               &   2nd moment         \\
\hline
Favored          & $D_u^{\pi^+}$          &  0.401$\pm$0.052     \\
Disfavored       & $D_{\bar u}^{\pi^+}$   &  0.094$\pm$0.029     \\
\hline
Favored          & $D_u^{K^+}$            &  0.0740$\pm$0.0268   \\
Favored          & $D_{\bar s}^{K^+}$     &  0.0878$\pm$0.0506   \\
Disfavored       & $D_{\bar u}^{K^+}$     &  0.0255$\pm$0.0173   \\
\hline
Favored          & $D_u^{p}$  \  \        &  0.0732$\pm$0.0113   \\
Disfavored       & $D_{\bar u}^{p}$ \ \   &  0.0084$\pm$0.0057   \\
\hline
\end{tabular}
\end{center}
\end{wraptable}

From the analyses of ordinary hadrons, pion, kaon, and proton,
we found characteristic differences between the favored and
disfavored FFs. In Table \ref{tab:2nd-moments},
the second moments are shown for $\pi^+$, $K^+$, and $p$. 
It is clear that the moments are larger for the favored functions
than the ones for the disfavored functions. It indicates that
internal quark configuration can be found by looking at 
flavor dependence of the FFs. This fact suggested us 
to use the FFs for exotic hadron search \cite{hkos08}.

The exotic means a hadron with internal quark configuration
other than ordinary $q\bar q$ and $qqq$. This topic has been
investigated for a long time; however, an undoubted evidence
has not been found yet. In the last several years, there have
been reports on exotic candidates mainly from Belle and BaBar
collaborations in charmed hadrons. In our work, we investigated
a possibility that the FFs can be used for an exotic hadron
search by using differences between the favored and disfavored FFs.

As one of the exotic mesons, we investigated a possibility of
determining quark configuration of $f_0 (980)$, which 
structure has been controversial for many years.
According to a simple quark model, it is described
by the configuration $(u\bar u + d\bar d)/\sqrt{2}$.
However, it is known that theoretical strong decay widths are 
an order of magnitude larger than the experimental width \cite{f0-decay}. 
Therefore, it is considered to be $s\bar s$, tetraquark, or
$K \bar K$ molecule state. It used to be considered as
a glueball candidate, but recent lattice QCD estimates
indicate that the lowest scalar-meson mass is about
1700 MeV. There is an indication that the internal configuration
could be determined by the radiative decay 
$\phi \rightarrow f_0 \gamma$ and 2$\gamma$ decay 
$f_0 \rightarrow 2\gamma$ \cite{hkos08,f0-e1}; however, 
the FF method could become a better way for judging
its internal structure as well as other exotic-meson
configurations.

\begin{table}[t]
\caption{Possible $f_0(980)$ configurations and their features 
         in FFs at small $Q^2$ \cite{hkos08}.}
\label{tab:f0-config}
\begin{center}
\begin{tabular}{cccc}
\hline
Type                   & Configuration 
                       & Second moments
                       & Peak positions      \\
\hline
Nonstrange $q\bar q$   & $(u\bar u+d\bar d)/\sqrt{2}$  
                       & $M_s<M_u<M_g$
                       & $z_u^{\rm max}>z_s^{\rm max}$   \\  
Strange    $q\bar q$   & $s\bar s$                 
                       & $M_u  <   M_s \lesssim M_g$    
                       & $z_u^{\rm max}<z_s^{\rm max}$   \\  
Tetraquark (or $K\bar K$) & $(u\bar u s\bar s+d\bar d s\bar s)/\sqrt{2}$  
                       & $M_u \sim M_s \lesssim M_g$
                       & $z_u^{\rm max} \sim z_s^{\rm max}$   \\  
Glueball               & $gg$ 
                       & $M_u \sim M_s < M_g$
                       & $z_u^{\rm max} \sim z_s^{\rm max}$   \\  
\hline
\end{tabular}
\end{center}
\end{table}

We summarized in Table \ref{tab:f0-config} how to judge 
the structure of the $f_0$ meson, especially
by the second moments and $z$-dependent functional forms.
All the possible configurations are considered in the table.
For example, if $f_0$ is an $s\bar s$ state, $s\bar s$ is formed
from $s$ by creating an $s\bar s$ pair from a radiated gluon.
In the same way, a color neutral $s\bar s$ can
be formed from a gluon by its splitting into an $s\bar s$ pair
and a subsequent gluon radiation for color neutrality
(see Ref.9 for details).
The gluon can be radiated from either $s$ or $\bar s$, so that
$M_g$ could be larger than $M_s$ as indicated in the table.
If $s$ is a favored quark, a significant portion of its
energy (namely large $z$) is transferred to $f_0 (s\bar s)$. 
It leads to a functional form which is mainly distributed in
the large-$z$ region. In the table, this is denoted
as $z_u^{max} < z_s^{max}$.
In the same way, the relations in the second moments
and the functional forms are listed for other configurations.
Our suggestions are intended to give an idea on the criteria,
and further details need to be investigated in more sophisticated 
hadron models.

There are data on the FFs of $f_0$ in the $e^+ e^-$ annihilation. 
We have done a global analysis for determining the FFs in the same way
with the analyses in Sec. \ref{method} \cite{hkos08}. 
This is intended to judge the quark configuration
of $f_0$ by using the criteria of Table \ref{tab:f0-config}.
The determined second moments are given by
$M_u  = 0.0012 \pm 0.0107$,
$M_s  = 0.0027 \pm 0.0183$, and
$M_g  = 0.0090 \pm 0.0046$.
Then, the moment ratio becomes
$M_u/M_s=0.43\pm 6.73$. From the ratio $0.43$, the $f_0$ seems to
be mainly an $s\bar s$ state. However, it is obvious from its huge error $6.73$
that a clear determination is currently not possible.

\begin{wrapfigure}{r}{0.40\textwidth}
   \vspace{-0.4cm}
   \begin{center}
       \epsfig{file=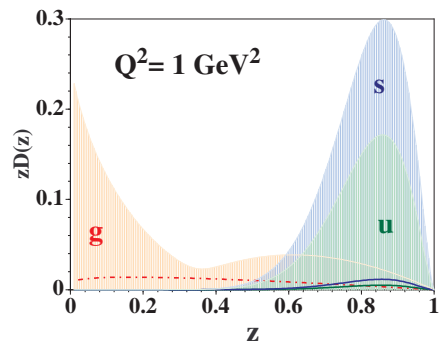,width=0.40\textwidth} \\
   \end{center}
   \vspace{-0.1cm}
\caption{Determined FFs for $f_0(980)$ \cite{hkos08}.
Note that uncertainty bands are much larger than the FFs.}
\label{fig:f0-ffs}
\end{wrapfigure}

The obtained FFs from the global analysis are shown in Fig. \ref{fig:f0-ffs}.
The up- and strange-quark functions are distributed mainly in the large-$z$
region, which may indicate that $u$ and $s$ could be important constituents
of $f_0$. However, the uncertainties of the determined FFs are huge and 
they are an order of magnitude larger than the FFs themselves. 
Therefore, it is not possible to draw a conclusion on its structure 
from the global analysis at this stage.
Hopefully, much better data will be reported in the near future possibly
from the Belle collaboration \cite{belle} for determining
structure of exotic hadrons including $f_0 (980)$.

\section{Summary}
\label{summary}

Our studies on the fragmentation functions were reported.
First, global analyses have been done for pion, kaon, 
and proton for determining their FFs. 
In the past, it was not clear why there are large discrepancies
among various parametrizations on disfavored-quark and gluon
FFs. We clarified that they are consistent with each other
by estimating the uncertainties of the FFs by the Hessian method
in the sense that all the distributions are within our error bands.
We also clarified the role of NLO terms in reducing the uncertainties
by comparing the determined FFs and their uncertainties in the LO and NLO. 
Our code for calculating the FFs was supplied on our web site \cite{ffs-web}.

Next, we investigated a possibility of using the FFs for finding
internal structure of exotic hadrons by using differences
between the favored and disfavored FFs. We proposed to use the second moments
and $z$-dependent functional forms of the FFs for determining the internal
structure. As an example, the $f_0 (980)$ meson was studied by 
considering all the possibilities of $q\bar q$, tetraquark, and glueball
configurations. A global analysis has been done also for the FFs of
$f_0$ by using the current $e^+ e^-$ data; however, they were not accurately 
determined to the level of finding the internal structure, namely a difference
between the favored-quark and disfavored-quark (and gluon) FFs. 
Much more accurate data are needed to discuss the internal structure
by the FFs.

\section*{Acknowledgements}
The authors would like to thank discussion with T. Nagai, M. Oka,
and K. Sudoh on the fragmentation functions. 



\end{document}